\documentclass[10pt,twocolumn,letterpaper]{article}

\usepackage[pagenumbers]{cvpr}

\usepackage{subcaption}
\usepackage{graphicx}
\usepackage{amsmath}
\usepackage{amssymb}
\usepackage{booktabs}
\usepackage{multirow}

\usepackage{ifsym}
\usepackage{color}
\usepackage{xcolor}
\definecolor{citecolor}{HTML}{0071bc}
\usepackage[pagebackref,breaklinks,colorlinks,citecolor=citecolor,bookmarks=false]{hyperref}

\usepackage[capitalize]{cleveref}
\crefname{section}{Sec.}{Secs.}
\Crefname{section}{Section}{Sections}
\Crefname{table}{Table}{Tables}
\crefname{table}{Tab.}{Tabs.}

\def\cvprPaperID{8888}
\def\confName{CVPR}
\def\confYear{2022}

\makeatletter
\def\thanks#1{\protected@xdef\@thanks{\@thanks
        \protect\footnotetext{#1}}}
\makeatother

\begin{document}

\title{VideoINR: Learning Video Implicit Neural Representation for\\Continuous Space-Time Super-Resolution}

\author{
    Zeyuan Chen$^{1}$ ~~~~
    Yinbo Chen$^{2}$~~~~
    Jingwen Liu$^{2}$~~~~
    Xingqian Xu$^{3,6}$~~~~
    Vidit Goel$^6$~~~~\\
    Zhangyang Wang$^5$~~~~
    Humphrey Shi$^{6,5,3\dagger}$~~~~
    Xiaolong Wang$^{2\dagger}$~~~~~\thanks{$^\dagger$ Corresponding authors.}
    \\
    $^1$\normalsize USTC~~~~$^2$\normalsize UC San Diego~~~~
    $^3$\normalsize UIUC~~~~$^4$\normalsize UT Austin~~~~$^5$\normalsize U of Oregon~~~~$^6$\normalsize Picsart AI Research (PAIR)\\
}
\maketitle

\begin{abstract}
Videos typically record the streaming and continuous visual data as discrete consecutive frames. Since the storage cost is expensive for videos of high fidelity, most of them are stored in a relatively low resolution and frame rate. Recent works of Space-Time Video Super-Resolution (STVSR) are developed to incorporate temporal interpolation and spatial super-resolution in a unified framework. However, most of them only support a fixed up-sampling scale, which limits their flexibility and applications. In this work, instead of following the discrete representations, we propose Video Implicit Neural Representation (\textbf{VideoINR}), and we show its applications for STVSR. The learned implicit neural representation can be decoded to videos of arbitrary spatial resolution and frame rate. We show that VideoINR achieves competitive performances with state-of-the-art STVSR methods on common up-sampling scales and significantly outperforms prior works on continuous and out-of-training-distribution scales. Our project page is at \href{http://zeyuan-chen.com/VideoINR/}{http://zeyuan-chen.com/VideoINR/}.
\end{abstract}

\section{Introduction}
\label{sec:intro}
We observe the visual world in the form of streaming and continuous data. However, when we record such data with a video camera in a computer, it is often stored with limited spatial resolutions and temporal frame rates. Because of the high cost on recording and storing large time-scales of video data, oftentimes our computer vision system will need to process low-resolution and low frame rate videos. This introduces challenges in recognition systems such as video object detection~\cite{zhu2017flow}, and we are still struggling at learning to recognize motion and actions from discrete frames~\cite{carreira2017quo,feichtenhofer2019slowfast}. When presenting the video back to humans (e.g., on a TV), it is essential to visualize it in high resolution and high frame rate for user experience. How to recover the low resolution video back to high resolution in space and time becomes an important problem and the first step for many downstream applications. 

\begin{figure}
    \centering
    \includegraphics[width=\linewidth]{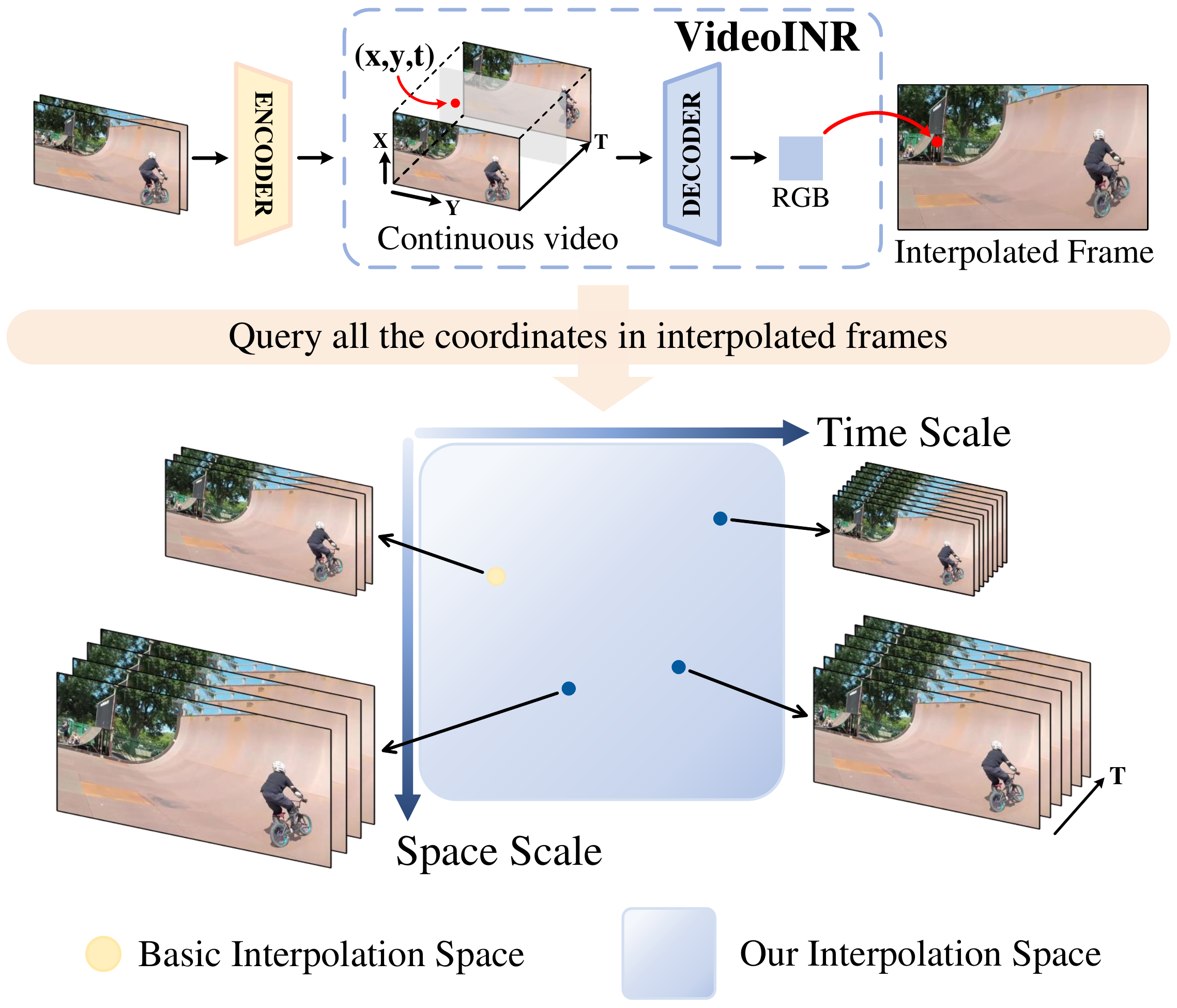}
    \caption{Video Implicit Neural Representation (VideoINR) maps any 3D space-time coordinate to an RGB value. This nature enables extending the interpolation space of STVSR from fixed space and time scales to arbitrary frame rate and spatial resolution.}
    \vspace{-1em}
    \label{fig:teaser}
\end{figure}

Space-Time Video Super-Resolution (STVSR) approaches~\cite{shechtman2002stvsr,mudenagudi2010stvsr,shahar2011stvsr,kim2020fisr,haris2020starnet,xiang2020zooming,xu2021tmnet} are developed to increase the spatial resolution and frame rate at the same time given a low-resolution and low frame rate video as the input. Instead of performing super-resolution in space and time separately in two stages, researchers recently propose to simultaneously perform super-resolution in one stage~\cite{kim2020fisr,haris2020starnet,xiang2020zooming,xu2021tmnet}. Intuitively, the aggregated information in time from multiple frames can reveal missing details for each frame when spatial scaling is applied, and the temporal interpolation can be more smooth and accurate given higher and richer spatial representation. The one-stage end-to-end training has shown to unify the benefits from both sides. While these results are encouraging, most approaches can only perform super-resolution to a fixed space and time scale ratio. 

In this paper, instead of super-resolution in a fixed scale, we propose to learn a continuous video representation, which allows to sample and interpolate the video frames in arbitrary frame rate and spatial resolution at the same time. Our key idea is to learn an implicit neural representation, which is a neural function that takes a space-time coordinate as input, and outputs the corresponding RGB value. Since we can sample the coordinate continuously, the video can be decoded in any spatial resolution and frame rate. Our work is inspired by recent progress on  implicit functions for 3D shape representations~\cite{deng2020nasa,genova2019learning,genova2020local,michalkiewicz2019implicit} and image representations with Local Implicit Image Functions (LIIF) using a ConvNet~\cite{chen2021liif}. Different from images, where interpolation in space can be based on the gradients between pixels, pixel gradients across frames with low frame rates are hard to compute. The network will need to understand the motion of the pixels and objects to perform interpolation, which could be hard to model by 2D or 3D convolutions alone. 

We propose a novel Video Implicit Neural Representation (VideoINR) as a continuous video representation. In the STVSR task, two low-resolution image frames are concatenated and forwarded to an encoder which generates a feature map with spatial dimensions. VideoINR then serves as a continuous video representation over the generated feature map. It first defines a spatial implicit neural representation for a continuous spatial feature domain, from which a high-resolution image feature is sampled according to all query coordinates. Instead of using convolutional operations to perform temporal interpolation, we learn a temporal implicit neural representation to first output a motion flow field given the high-resolution feature and the sampling time as inputs. This flow field will be applied back to warp the high-resolution feature which will be decoded to the target video frame. Since all the operations are differentiable, we can learn the motion in feature level end-to-end without any extra supervision besides the reconstruction error. To summarize, given the input frames, an encoder generates a feature map, which can be then decoded by VideoINR to arbitrary spatial resolution and frame rate.

In our experiments, we demonstrate that VideoINR can not only represent video in arbitrary space and time resolutions on the scales within the training distributions, but also extrapolate to out-of-distribution frame rates and spatial resolutions. Given the learned continuous function, instead of decoding the whole video each time, it allows the flexibility to decode only a certain region and time scale when needed. We conduct experiments with Vid4~\cite{liu2011vid4}, GoPro~\cite{nah2017gopro} and Adobe240~\cite{su2017adobe} datasets. We demonstrate that VideoINR achieves competitive performances with state-of-the-art STVSR methods on in-distribution spatial and temporal scales and significantly outperforms other methods on out-of-distribution scales.

We highlight our main contributions as follows:
\begin{itemize}
\vspace{-0.05in}
    \item We propose a novel Video Implicit Neural Representation as a continuous video representation. 
    \vspace{-0.1in}
    \item The proposed approach allows for representing videos in arbitrary space and time resolution efficiently.
    \vspace{-0.1in}
    \item VideoINR achieves out-of-distribution generalization in space-time video super-resolution and outperforms baselines by a large margin.
\end{itemize}

\section{Related Work}
\label{sec:relate-work}
\noindent\textbf{Implicit Neural Representation.}
Implicit neural representations have been demonstrated as compact yet powerful continuous representations for various tasks, including 3D reconstruction~\cite{deng2020nasa,genova2019learning,genova2020local,michalkiewicz2019implicit} and generation~\cite{schwarz2020graf,chan2021pi,devries2021unconstrained}. These representations typically represent signals as a neural function that maps coordinates to signed distance~\cite{Park_2019_CVPR}, occupancy~\cite{Mescheder_2019_CVPR,Chen_2019_CVPR}, or density and RGB values in a neural radiance field (NeRF~\cite{mildenhall2020nerf}). Recent works also show promising results of applying this idea for modeling 2D images~\cite{chen2021liif,xu2021ultrasr,skorokhodov2021adversarial,anokhin2021image,karras2021alias}. Our continuous video representation is inspired by this rapidly growing field and has specific designs for videos, where a learnable flow can exploit the correspondences in video frames with inductive bias.

\noindent\textbf{Video Frame Interpolation. }Video Frame Interpolation (VFI) aims to synthesize unseen frames between the input video frames. Meyer~\etal~\cite{meyer2015phasevfi} proposed a phase-based method where information across levels of a multi-scale pyramid is combined for the synthesis of interpolated frames. Niklaus~\etal~\cite{niklaus2017kernelvfi1,niklaus2017kernelvfi2} introduced a series of kernel-based VFI algorithms in which they took pixel synthesis for the target frame as local convolution over input frames. Optical flow based VFI methods~\cite{jiang2018super,niklaus2018kernelvfi3,xue2019vimeo,bao2019dain,xu2019qvi,niklaus2020softmax} utilized optical flow prediction networks (\eg PWC-Net~\cite{sun2018pwcnet}) to compute bidirectional flows between input frames, which served as the guidance for new frame synthesis. Additional information including occlusion masks~\cite{jiang2018super,xue2019vimeo}, depth maps~\cite{bao2019dain}, and cycle consistency~\cite{reda2019cyclevfi} were also incorporated in the models for better performances.

\begin{figure*}[t]
\vspace{-0.3em}
    \centering
    \includegraphics[width=\linewidth]{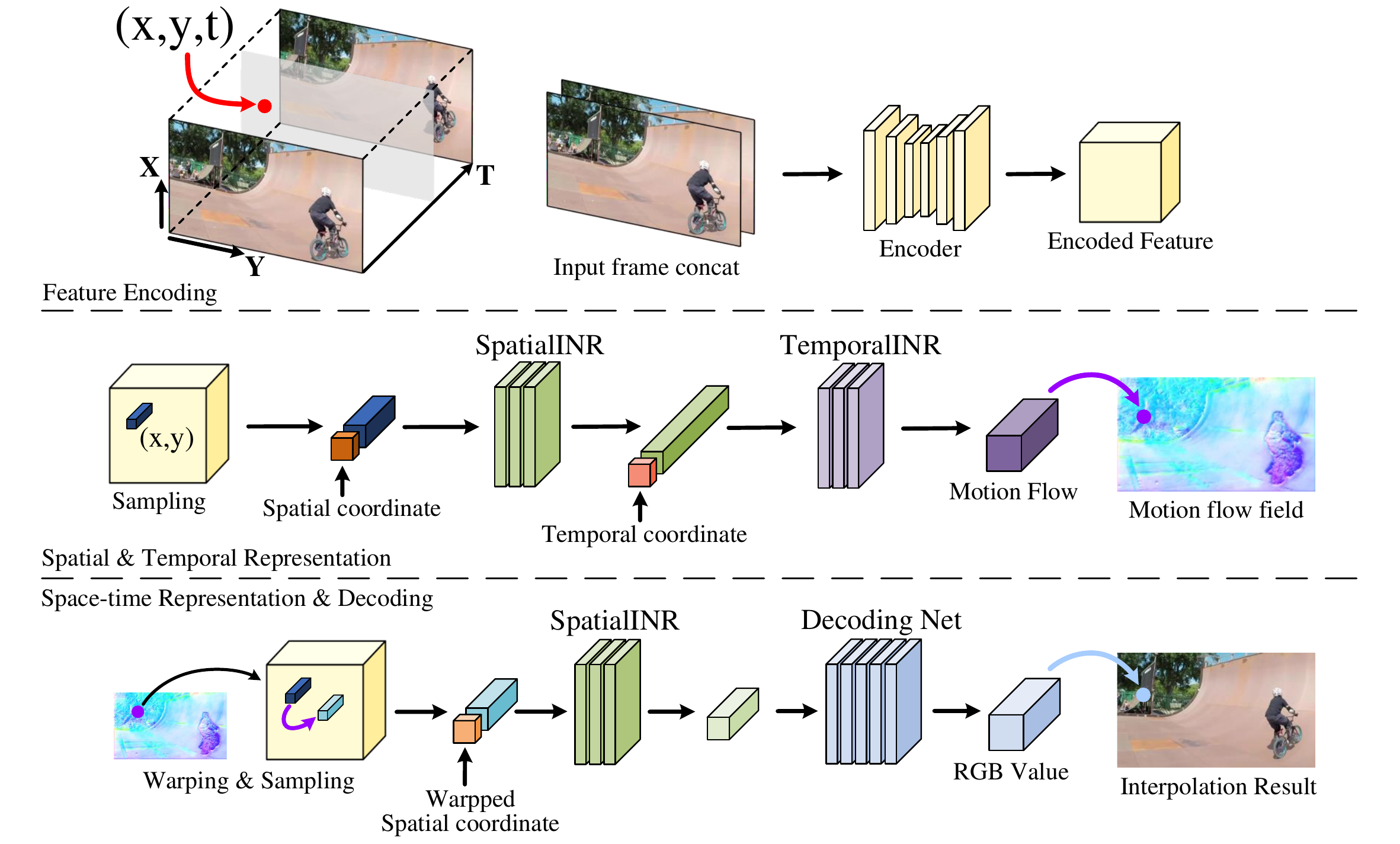}
    \vspace{-1.25em}
    \caption{\textbf{An overview of our Video Implicit Neural Representation (VideoINR).} Two consecutive input frames are concatenated and encoded as a discrete feature map. Based on the feature, the spatial and temporal implicit neural representations decode a 3D space-time coordinate to a motion flow vector. We then sample a new feature vector by warping according to the motion flow, and decode it as the RGB prediction of the query coordinate. We omit the multi-scale feature aggregation part in this figure.}
    \vspace{-0.75em}
    \label{fig:overview}
\end{figure*}

\noindent\textbf{Video Super-Resolution. }Video Super-Resolution (VSR) aims at increasing the spatial resolutions of low-resolution videos. Earlier approaches~\cite{tao2017vsr,caballero2017vsr2,xue2019vimeo} were typically built on the sliding-window framework, where they predicted optical flows between input frames and performed spatial warping for explicit feature alignment. Later on, implicit alignment started a new trend in this task~\cite{jo2018deepvsr,tian2020tdan,wang2019edvr,isobe2020videovsr,chan2021basicvsr}. For instance, TDAN~\cite{tian2020tdan} adopts deformable convolutions (DCNs)~\cite{dai2017deformable,zhu2019deformable} to align different input frames at feature levels. EDVR~\cite{wang2019edvr} further extends DCNs to a multi-scale fashion for more accurate alignment. Kelvin~\etal introduced BasicVSR~\cite{chan2021basicvsr}, in which they analyzed basic components for VSR models and suggested a bidirectional propagation scheme to maximize the gathered information from input. 

\noindent\textbf{Space-time Video Super-Resolution}
The target of Space-Time Video Super-Resolution (STVSR) is to simultaneously increase the spatial and temporal resolutions of the given low-resolution low frame rate videos. Shechtman~\etal~\cite{shechtman2002stvsr} tackled this problem by combining information from multiple input video sequences and applying a directional space-time regularization. Mudenagudi~\etal~\cite{mudenagudi2010stvsr} proposed a unified framework for STVSR in which videos are modeled as Markov random fields, and the maximum a posteriori estimates are taken as final solutions. Shahar~\etal~\cite{shahar2011stvsr} introduced an effective space-time patch recurrence prior for STVSR. Recently, with the advances in deep learning, researchers started to employ powerful convolutional neural networks to address the task~\cite{kim2020fisr,haris2020starnet,xiang2020zooming,xu2021tmnet}. Xiang~\etal~\cite{xiang2020zooming} proposed a unified neural network for synthesizing the feature of the missing frame and used a deformable ConvLSTM to align and aggregate extracted temporal information for reconstruction. STARNet~\cite{haris2020starnet} leveraged mutually informative relationships between time and space with the assistance of additional optical flow inputs. TMNet~\cite{xu2021tmnet} proposed a temporal modulation block to modulate deformable convolution kernels for supporting frame interpolation at arbitrary time instances. All these STVSR methods are designed to perform super-resolution on a specific up-sampling space scale defined before training, and some of them~\cite{haris2020starnet,xiang2020zooming} can only infer intermediate frames at pre-defined times. Therefore, the application scopes of these methods are limited. VideoINR serves as a continuous video representation that supports frame interpolation at arbitrary spatial resolution and frame rate. VideoINR is more flexible during the application and can be employed in more circumstances, such as non-uniform interpolation and video zoom-in in local regions.

\section{Video Implicit Neural Representation}
\label{sec:method}
Given a video with limited spatial resolution and frame rate, our goal is to find a continuous representation for the video. The representation interprets arbitrary space-time coordinate $(x_s,x_t)$ into RGB values. To this end, we introduce Video Implicit Neural Representation (VideoINR), which enables continuous space-time super-resolution. It is parameterized by multi-layer perceptrons (MLPs) and takes the form
\begin{equation}
    s = f(x_s, x_t)
\end{equation}
where $f$ is the proposed video representation defined by the encoded feature and network parameters. $x_s$ is the 2D spatial coordinate, $x_t$ is the temporal coordinate, and $s$ is the predicted RGB value. For learning such implicit neural representation, we propose to decouple space and time and learn a continuous representation for each of them.

Figure~\ref{fig:overview} illustrates an overview of our model. Given a space-time coordinate $(x_s,x_t)$ and the feature extracted from input frames by an encoder, the Spatial Implicit Neural Representation (SpatialINR) decodes the spatial coordinate $x_s$ and output a corresponding feature vector (Sec.~\ref{sec:SIF}). The feature is then forwarded to the Temporal Implicit Neural Representation (TemporalINR) for the motion flow at the query coordinate (Sec.~\ref{sec:TIF}). The flow is applied back to warp the continuous feature defined by SpatialINR for a new feature vector (Sec.~\ref{sec:STIF}) which is finally decoded to the target RGB value (Sec.~\ref{sec:decoding}).

\subsection{Continuous Spatial Representation}
\label{sec:SIF}
Inspired by LIIF~\cite{chen2021liif}, we learn a Spatial Implicit Neural Representation (SpatialINR) that defines a continuous 2D feature domain by the discrete encoded feature map. This continuous domain decodes arbitrary 2D spatial coordinate into a corresponding feature vector. Specifically, the feature vectors generated by the encoder are evenly distributed in the 2D space. We sample the feature vector (the dark blue cuboid in Fig~\ref{fig:overview}) nearest to the queried spatial coordinate $x_s$, concatenate it with the
relative position information between query coordinate and feature vector, and input them into the a function $f_s$ to output the continuous feature at $x_s$ (the green cuboid in Fig~\ref{fig:overview}). This process could be expressed as
\begin{equation}
    \mathcal{F}_s(x_s) = f_s(z^*, x_s-v^*)
\label{eq:sif}
\end{equation}
where $\mathcal{F}_s$ is the continuous feature domain defined by SpatialINR, $z^*$ is the feature vector nearest to the query coordinate $x_s$ and $v^*$ is the spatial coordinate of the feature vector $z^*$. 

The main difference between LIIF and SpatialINR is that LIIF is proposed for continuous image representation, while SpatialINR defines a continuous feature domain, which is supposed to be further utilized for modeling temporal information in videos.

\subsection{Continuous Temporal Representation}
\label{sec:TIF}
The proposed SpatialINR defines a new continuous feature domain in 2D space. Our next step is to learn the continuous Temporal Implicit Neural Representation (TemporalINR) and extend the feature domain from 2D space to 3D space and time, which can be achieved by decoding the temporal coordinate $x_t$. Directly generating the target decoded feature by a network can be fairly difficult, as the network has to learn not only the motion patterns between input frames but also the context information. Instead, we propose to learn a continuous motion flow field for the continuous temporal representation.

Particularly, given a space-time coordinate $(x_s,x_t)$ and two consecutive input frames $I_0$ and $I_1$, TemporalINR maps the coordinate to a motion flow
\begin{equation}
    \mathcal{M}(x_s,x_t) = f_t(x_s, x_t, I_0, I_1)
\label{eq:tif1}
\end{equation}
where $\mathcal{M}$ is the continuous motion flow field and $f_t$ is the function for TemporalINR. Benefiting from the 2D continuous feature domain provided by SpatialINR, we could replace $I_0$, $I_1$, and $x_s$ by the continuous feature at $x_s$. Thus the equation can be written as
\begin{equation}
    \mathcal{M}(x_s,x_t) = f_t(x_t, \mathcal{F}_s(x_s))
\end{equation}
where $\mathcal{F}_s(x_s)$ is the feature domain defined in Eq~\ref{eq:sif}.

\subsection{Space-Time Continuous Representation}
\label{sec:STIF}
With two continuous representations for space and time, we aim at combining them into a unified space-time continuous representation for videos. Starting from a space-time coordinate $(x_s,x_t)$, we first use SpatialINR to predict the continuous feature at $x_s$. TemporalINR is then utilized for generating the motion flow of the query coordinate. Based on these outputs, we obtain the space-time feature by warping the continuous feature domain. The warped feature at $x_s$ corresponds to the continuous feature at $x_s'$. The relationship between two coordinates can be written as 
\begin{equation}
    x_s' = x_s+\mathcal{M}(x_s,x_t)
\end{equation}
where $\mathcal{M}(x_s,x_t)$ is the motion flow vector at $(x_s,x_t)$.

We query this new spatial coordinate in the continuous 2D feature domain and obtain a new feature vector (the light green cuboid in Fig~\ref{fig:overview}), which is treated as the feature of our continuous space-time representation at coordinate $(x_s,x_t)$. Accordingly, the continuous space-time feature $\mathcal{F}_{st}$ can be formulated as
\begin{equation}
    \mathcal{F}_{st}(x_s,x_t)=\mathcal{F}_s(x_s')=\mathcal{F}_s(x_s+\mathcal{M}(x_s,x_t))
\end{equation}

In practice, we generate two independent flows for the motion flow field, and concatenate corresponding warped features. Intuitively, TemporalINR may implicitly learn bi-directional correspondences between the target frame and input frames, without explicit supervision. 
                                                                   
\subsection{Feature Decoding}
\label{sec:decoding}
Based on the continuous space-time representation, we can get the feature corresponding to any space-time coordinate. The final step is to decode the feature as an RGB value. A straightforward design is to take the obtained space-time feature for decoding directly. However, due to the MLP-based network architecture, the RGB value of every predicted pixel depends on a single feature vector, leading to a limited size of the network receptive field. To alleviate the negative impact of this disadvantage, we enrich the input information of the decoding network by aggregating features of different scales. In detail, we incorporate the encoded feature as well as  two input frames for decoding. Since these additional features are typically of low-resolution compared with the target resolution, we sample feature vectors corresponding to the query coordinate by bilinear interpolation. All features are then combined together for predicting the RGB output. 

\subsection{Frame synthesis}
\label{sec:synthesis}
From Section~\ref{sec:SIF} to~\ref{sec:decoding}, we focus on predicting the RGB value at a specific coordinate. To synthesize an entire frame, we need to query coordinates of all pixels of it. Given these coordinates, we can convert the continuous feature from SpatialINR into a high-resolution feature map. We can also generate a whole motion flow field for the latent high-resolution interpolated frame. Therefore, we do not have to forward SpatialINR twice before and after warping as in the situation of one input coordinate. Instead, we directly warp the whole high-resolution feature map based on the motion flow and input the warped feature into the decoding network to synthesize the target frame at one time.

\section{Experiments}
\subsection{Experimental Setup}
\label{sec:expset}
\noindent\textbf{Dataset.} We use Adobe240 dataset~\cite{su2017adobe} as the training set, which includes 133 videos in 720P taken by hand-held cameras. We follow~\cite{xu2021tmnet} to split these videos into the train, validation, and test subsets with 100, 16, and 17 videos. All videos are converted into image sequences for training and testing. Each sequence contains approximately 3000 frames which are treated as high-resolution frames in training. The low-resolution counterparts are then generated by imresize function in Matlab with the default setting of bicubic interpolation. We use a sliding window to select frames from the image sequences for training. The length of the sliding window is set to 9. We take the $1^{st}$ and $9^{th}$ frames as network inputs. The $2^{nd}$ to $7^{th}$ frames serve as ground-truth frames, and we randomly select three of them as the supervision of our network in every iteration. VideoINR is trained by two stages. In the first stage, we fixed the down-sampling space scale to $\times4$. In the second stage, we randomly sample scales in a uniform distribution $\mathcal{U}(1,4)$. We provide more discussion about this two-stage training strategy in Section~\ref{sec:ablation}.

\begin{table*}[h]
\caption{\textbf{Quantitative comparison on benchmark datasets }including Vid4~\cite{liu2011vid4}, GoPro~\cite{nah2017gopro} and Adobe240~\cite{su2017adobe}.  The best three results are highlighted in \textcolor{red}{red}, \textcolor{blue}{blue}, and \textbf{bold}. We omit the results of Zooming SlowMo and VideoINR-\textit{Fixed} on GoPro-\textit{Average} and Adobe240-\textit{Average} as the two models are trained for synthesizing frames only at fixed times.}
\resizebox{\textwidth}{!}{
\begin{tabular}{cc|ccccccccccc}
\hline
\multirow{2}{*}{\begin{tabular}[c]{@{}c@{}}VFI\\Method\end{tabular}} & \multirow{2}{*}{\begin{tabular}[c]{@{}c@{}}SR\\Method\end{tabular}} & \multicolumn{2}{c}{Vid4} & \multicolumn{2}{c}{GoPro-\textit{Center}} & \multicolumn{2}{c}{GoPro-\textit{Average}} & \multicolumn{2}{c}{Adobe-\textit{Center}} & \multicolumn{2}{c}{Adobe-\textit{Average}} & \multirow{2}{*}{\begin{tabular}[c]{@{}c@{}}Parameters\\   (Million)\end{tabular}} \\ 
 & & PSNR & SSIM & PSNR & SSIM & PSNR & SSIM & PSNR & SSIM & PSNR & SSIM & \\ 
\hline \hline
SuperSloMo~\cite{jiang2018super} & Bicubic & 22.42 & 0.5645 & 27.04 & 0.7937 & 26.06 & 0.7720 & 26.09 & 0.7435 & 25.29 & 0.7279 & 19.8 \\
SuperSloMo~\cite{jiang2018super} & EDVR~\cite{wang2019edvr} & 23.01 & 0.6136 & 28.24 & 0.8322 & 26.30 & 0.7960 & 27.25 & 0.7972 & 25.95 & 0.7682 & 19.8+20.7 \\ 
SuperSloMo~\cite{jiang2018super} & BasicVSR~\cite{chan2021basicvsr} & 23.17 & 0.6159 & 28.23 & 0.8308 & 26.36 & \textbf{0.7977} & 27.28 & 0.7961 & 25.94 & 0.7679 & 19.8+6.3 \\
\hline
QVI~\cite{jiang2018super} & Bicubic & 22.11 & 0.5498 & 26.50 & 0.7791 & 25.41 & 0.7554 & 25.57 & 0.7324 & 24.72 & 0.7114 & 29.2 \\
QVI~\cite{jiang2018super} & EDVR~\cite{wang2019edvr} & 23.60 & 0.6471 & 27.43 & 0.8081 & 25.55 & 0.7739 & 26.40 & 0.7692 & 25.09 & 0.7406 & 29.2+20.7 \\ 
QVI~\cite{jiang2018super} & BasicVSR~\cite{chan2021basicvsr} & 23.15 & 0.6428 & 27.44 & 0.8070 & 26.27 & 0.7955 & 26.43 & 0.7682 & 25.20 & 0.7421 & 29.2+6.3 \\
\hline
DAIN~\cite{bao2019dain} & Bicubic & 22.57 & 0.5732 & 26.92 & 0.7911 & 26.11 & 0.7740 & 26.01 & 0.7461 & 25.40 & 0.7321 & 24.0 \\
DAIN~\cite{bao2019dain} & EDVR~\cite{wang2019edvr} & 23.48 & 0.6547 & 28.01 & 0.8239 & 26.37 & 0.7964 & 27.06 & 0.7895 & 26.01 & 0.7703 & 24.0+20.7 \\ 
DAIN~\cite{bao2019dain} & BasicVSR~\cite{chan2021basicvsr} & 23.43 & 0.6514 & 28.00 & 0.8227 & \textbf{26.46} & 0.7966 & 27.07 & 0.7890 & \textbf{26.23} & \textbf{0.7725} & 24.0+6.3 \\
\hline
\multicolumn{2}{c|}{Zooming SlowMo~\cite{xiang2020zooming}} & \textbf{25.72} & \textbf{0.7717} & \textcolor{blue}{30.69} & \textcolor{blue}{0.8847} & - & - & \textcolor{red}{30.26} & \textcolor{red}{0.8821} & - & - & \textcolor{red}{11.10} \\ 
\multicolumn{2}{c|}{TMNet~\cite{xu2021tmnet}} & \textcolor{red}{25.96} & \textcolor{red}{0.7803} & 30.14 & 0.8692 & \textcolor{blue}{28.83} & \textcolor{blue}{0.8514} & 29.41 & 0.8524 & \textcolor{blue}{28.30} & \textcolor{blue}{0.8354} & \textbf{12.26} \\ 
\hline
\multicolumn{2}{c|}{VideoINR-\textit{fixed}} & \textcolor{blue}{25.78} & \textcolor{blue}{0.7730} & \textcolor{red}{30.73} & \textcolor{red}{0.8850} & - & - & \textcolor{blue}{30.21} & \textcolor{blue}{0.8805} & - & - & \textcolor{blue}{11.31} \\ 
    \multicolumn{2}{c|}{VideoINR} & 25.61 & 0.7709 & \textbf{30.26} & \textbf{0.8792} & \textcolor{red}{29.41} & \textcolor{red}{0.8669} & \textbf{29.92} & \textbf{0.8746} & \textcolor{red}{29.27} & \textcolor{red}{0.8651} & \textcolor{blue}{11.31} \\ 
\hline
\end{tabular}
}
\label{tab:sota-result}
\end{table*}

\begin{table*}[h]
\caption{\textbf{Quantitative comparison for out-of-distribution scales} on GoPro dataset. Model performances are evaluated by PSNR and SSIM. Some results of TMNet are bolded as it does not support generalizing to out-of-training-distribution space scales.}
\centering
\begin{tabular}{cc|cccc}
\hline
Time Scale & Space Scale & SuperSloMo~\cite{jiang2018super} + LIIF~\cite{chen2021liif} & DAIN~\cite{bao2019dain} + LIIF~\cite{chen2021liif} & TMNet~\cite{xu2021tmnet} & VideoINR \\ 
\hline \hline
$\times$6 & $\times$4 & 26.70 / 0.7988 & 26.71 / 0.7998 & 30.49 / 0.8861 & \textbf{30.78} / \textbf{0.8954} \\
$\times$6 & $\times$6 & 23.47 / 0.6931 & 23.36 / 0.6902 & - & \textbf{25.56} / \textbf{0.7671} \\
$\times$6 & $\times$12 & 21.92 / 0.6495 & 22.01 / 0.6499 & - & \textbf{24.02} / \textbf{0.6900} \\
\hline
$\times$12 & $\times$4 & 25.07 / 0.7491 & 25.14 / 0.7497 & 26.38 / 0.7931 & \textbf{27.32} / \textbf{0.8141} \\
$\times$12 & $\times$6 & 22.91 / 0.6783 & 22.92 / 0.6785 & - & \textbf{24.68} / \textbf{0.7358} \\
$\times$12 & $\times$12 & 21.61 / 0.6457 & 21.78 / 0.6473 & - & \textbf{23.70} / \textbf{0.6830} \\ 
\hline
$\times$16 & $\times$4 & 24.42 / 0.7296 & 24.20 / 0.7244 & 24.72 / 0.7526 & \textbf{25.81} / \textbf{0.7739} \\
$\times$16 & $\times$6 & 23.28 / 0.6883 & 22.80 / 0.6722 & - & \textbf{23.86} / \textbf{0.7123} \\
$\times$16 & $\times$12 & 21.80 / 0.6481 & 22.22 / 0.6420 & - & \textbf{22.88} / \textbf{0.6659} \\ 
\hline
\end{tabular}
\label{tab:ood}
\vspace{-3mm}
\end{table*}

Datasets including Vid4~\cite{liu2011vid4}, Adobe240~\cite{su2017adobe}, and GoPro~\cite{nah2017gopro} are used for evaluation. On Vid4, we only conduct experiments on single frame interpolation of STVSR. For Adobe240 and GoPro, we evaluate on their test set. The image sequences extracted from videos in the datasets are split into groups of 9-frame video clips. We feed the $1^{st}$ and $9^{th}$ frames down-sampled by scale $\times4$ in each clip into models to generate 9 high-resolution frames from $1^{st}$ to $9^{th}$. We separately evaluate the average metrics of the \textit{center} frames (\ie the $1^{st}$, $4^{th}$, $9^{th}$ frames) and all 9 output frames. They are denoted as -\textit{Center} and -\textit{Average} in Table~\ref{tab:sota-result}.

\noindent\textbf{Implementation details.} We use Adam optimizer~\cite{kingma2014adam} with $\beta_1=0.9$ and $\beta_2$=0.999. The learning rate is initialized as $1\times10^{-4}$ and is decayed to $1\times10^{-7}$ with a cosine annealing for every 150,000 iterations. The model is trained in a total of 600,000 iterations with batch size 24. The first training stage includes 450,000 iterations while the second stage includes 150,000 iterations. The input frames in one batch are down-sampled by the same space scale and randomly cropped into patches with size 32$\times$32. We perform data augmentation by randomly rotating $90^\circ$, $180^\circ$ and $270^\circ$, and horizontal-flipping. We use Zooming SlowMo~\cite{xiang2020zooming} as the encoder. For the two functions incorporated in continuous space and time representations, we utilize two 3-layer SIRENs~\cite{sitzmann2020siren} with hidden dimensions of $64,64,256$. For the decoding network, we employ a 4-layer SIREN with hidden dimensions of $64,64,256,256$. As suggested in~\cite{xiang2020zooming,xu2021tmnet}, we 
select the Charbonnier loss function for optimization.  

\noindent\textbf{Evaluation.} Peak-Signal-to-Noise Ratio (PSNR) and Structual Similarity Index (SSIM)~\cite{wang2004ssim} are employed to evaluate model performances. We also compare the model size and inference time to measure the efficiency of models.

\subsection{Comparison to State-of-the-arts}
\label{sec:sota}
We compare VideoINR with state-of-the-art two-stage and one-stage STVSR methods. For two-stage methods, we employ SuperSloMo~\cite{jiang2018super}, QVI~\cite{xu2019qvi}, and DAIN~\cite{bao2019dain} for video frame interpolation (VFI); Bicubic Interpolation, EDVR~\cite{wang2019edvr}, and BasicVSR~\cite{chan2021basicvsr} for video super-resolution (VSR). For one-stage methods, we compare VideoINR with recently developed Zooming SlowMo~\cite{xiang2020zooming} and TMNet~\cite{xu2021tmnet}. To perform fair comparisons, we train the three VFI methods and Zooming SlowMo from scratch on Adobe240 dataset. For TMNet, as mentioned in the original paper that a two-stage training scheme is needed for convergence, we pre-train the model on Vimeo90K~\cite{xue2019vimeo} dataset and fine-tune it on Adobe240 dataset~\cite{su2017adobe}. Therefore, TMNet is trained on more data compared with other methods, which may lead to some advantages in the comparison. To compare with Zooming SlowMo that only supports fixed frame interpolation, we train a new version of VideoINR named VideoINR-$\textit{fixed}$ of which the interpolation time is fixed to 0.5.

\begin{table}[t]
\caption{\textbf{Quantitative comparison of out-of-distribution performance between VideoINR and the baseline Zooming Slomo model~\cite{xiang2020zooming}}. Evaluated on GOPRO dataset. -$\times$A$\times$B refers to A up-sampling space scale and B up-sampling time scale.}
\centering
\begin{tabular}{c|cccccccc}
\hline
\multirow{2}{*}{\begin{tabular}[c]{@{}c@{}}Method\end{tabular}} &  \multicolumn{2}{c}{GoPro - $\times4\times$2} & \multicolumn{2}{c}{GoPro - $\times16\times$4} \\ 
 & PSNR & SSIM & PSNR & SSIM \\
\hline \hline
Zooming Slomo & 30.69 & 0.8847 & 23.38 & 0.6708 \\
VideoINR & 30.26 & 0.8792 & 23.45 & 0.6710  \\
\hline
\end{tabular}
\label{tab:zoomood}
\vspace{-5mm}
\end{table}

\begin{figure}
    \centering
    \includegraphics[width=\linewidth]{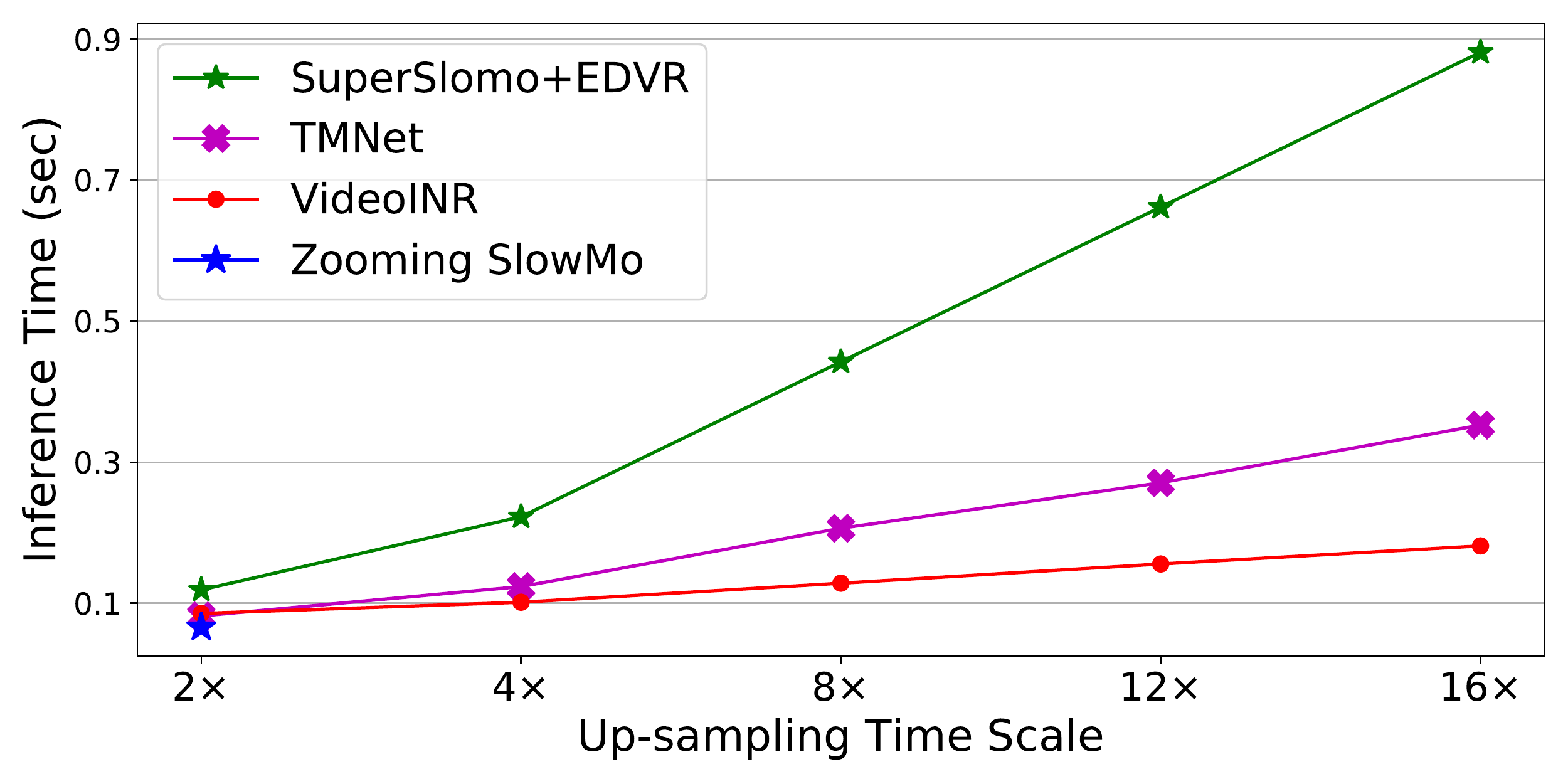}
    \vspace{-1.8em}
    \caption{\textbf{Inference time of STVSR models on different up-sampling time scales.} Space scale is set to 4. We select the most efficient two-stage method (SuperSlomo + EDVR) as a baseline.}
    \vspace{-1.5em}
    \label{fig:time}
\end{figure}

\begin{figure*}
    \centering
    \vspace{-0.5em}
    \includegraphics[width=1.03\linewidth]{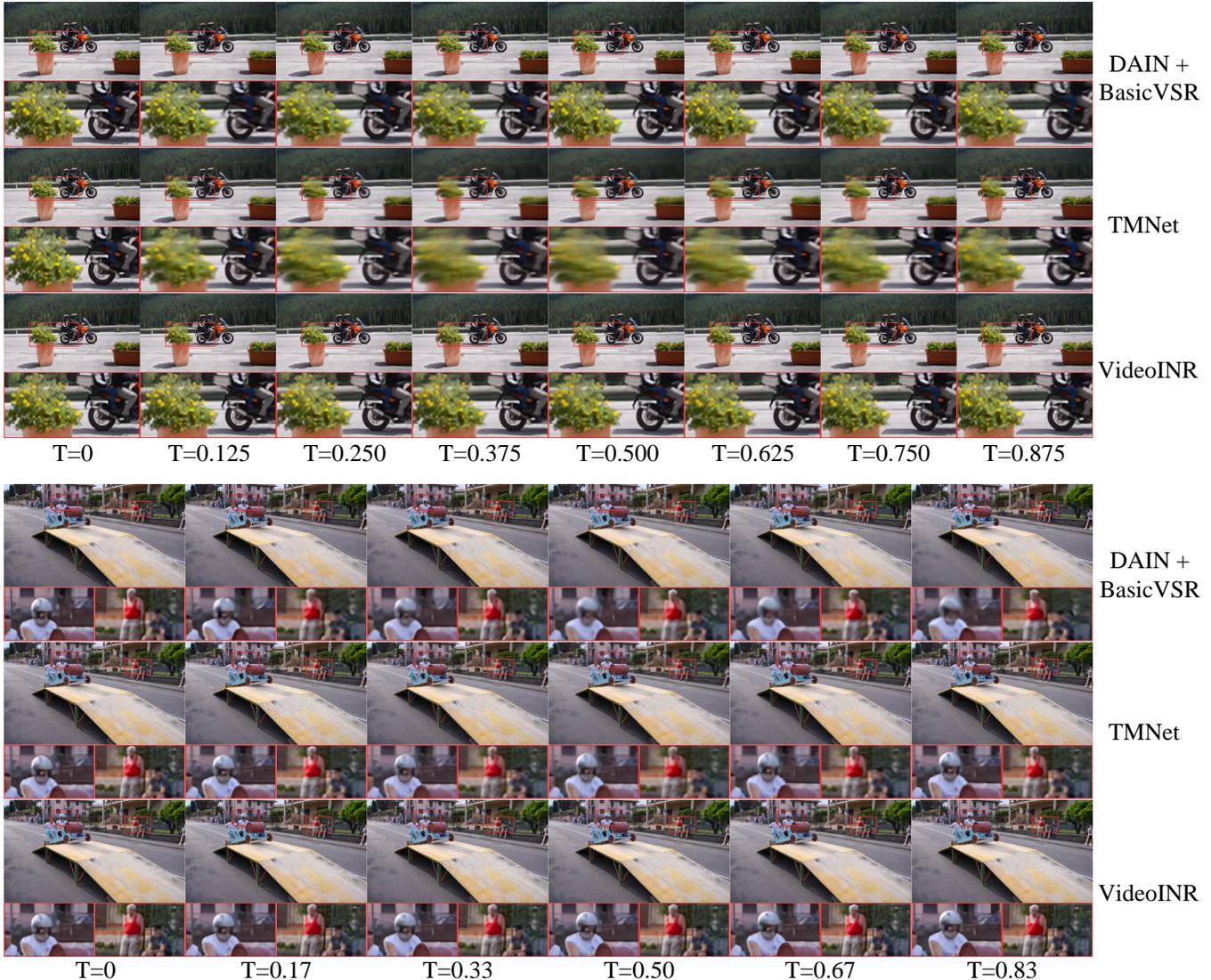}
    \vspace{-2em}
    \caption{\textbf{Qualitative comparisons of different STVSR methods on arbitrary frame interpolation.} The interpolation times of the first example are in the training distribution and the times of the second example are out-of-distribution. Best zoom in for better visualization.}
    \vspace{-1em}
    \label{fig:multiframe}
\end{figure*}

\noindent\textbf{Quantitative results.} We present in-distribution quantitative comparisons between VideoINR and other STVSR methods in Table~\ref{tab:sota-result}. On single frame interpolation of STVSR including Vid4, GoPro-$\textit{Center}$, and Adobe-$\textit{Center}$, VideoINR-$\textit{Fixed}$ achieves competitive performance compared with other state-of-the-art models, while the performance of VideoINR slightly suffers. We attribute this observation to the difference of training targets between VideoINR and VideoINR-$\textit{Fixed}$. The training settings of VideoINR-$\textit{Fixed}$ aim for synthesizing frames at pre-defined times. Therefore, it only learns fixed patterns between input frames instead of learning a continuous representation as VideoINR does, leading to advantages in performances. On Vid4, TMNet performs the best, and we assume this is because TMNet is trained with more data as we noted in Section~\ref{sec:sota}. For multiple frame interpolation of STVSR including GoPro-$\textit{Average}$ and Adobe-$\textit{Average}$, VideoINR achieves the best performance, which indicates that the proposed implicit neural representation provides advances on modeling the temporal information in videos. 

In Table~\ref{tab:ood}, we present comparisons of STVSR methods on out-of-distribution space and time scales. For two-stage STVSR methods, we select SuperSloMo and DAIN as VFI methods, and LIIF as the SR method since it can perform super-resolution on arbitrary up-sampling scales. We also take TMNet into the comparison as it could generalize on time scales. We produce experiments on GoPro~\cite{nah2017gopro} dataset. We observe that VideoINR outperforms other methods by a large margin, demonstrating the advantage of our continuous video representation in out-of-distribution generalization. In addition, we further compare VideoINR with Zooming SlowMo (the encoder for VideoINR) in out-of-distribution scales. As Zooming SlowMo only supports interpolating fixed frames, we apply the model twice to achieve out-of-distribution inferences. In Table~\ref{tab:zoomood}, we observe that while Zooming SlowMo performs slightly better on single frame interpolation ($\times4\times2$), VideoINR achieves better performance in out-of-distribution testing ($\times16\times4$).

We compare the inference time of STVSR methods in Figure~\ref{fig:time}. We observe that the efficiency of different methods is close at up-sampling time scale $\times2$, and VideoINR inferences faster than other models on multi-frame interpolation. We attribute this feature to the design of VideoINR, where all the latent frames between two input frames can be directly synthesized by MLPs after encoding.

\begin{table*}[h]
\vspace{-0.75mm}
\caption{\textbf{Ablation study on architecture designs of VideoINR.} Evaluated on GOPRO and Adobe240 dataset. -f/m refers to removing flow correspondence and multi-scale feature aggregation. -s refers to decoding both time and space by a single network.}
\vspace{-0.75mm}
\centering
\begin{tabular}{c|cccccccc}
\hline
\multirow{2}{*}{\begin{tabular}[c]{@{}c@{}}Architecture\\Design\end{tabular}} &  \multicolumn{2}{c}{GoPro-\textit{Center}} & \multicolumn{2}{c}{GoPro-\textit{Average}} &  \multicolumn{2}{c}{Adobe-\textit{Center}} & \multicolumn{2}{c}{Adobe-\textit{Average}} \\ 
 & PSNR & SSIM & PSNR & SSIM & PSNR & SSIM & PSNR & SSIM \\
\hline \hline
VideoINR & 30.26 & 0.8792 & 29.41 & 0.8669 & 29.92 & 0.8746 & 29.27 & 0.8651 \\
VideoINR (-f) & 29.63 & 0.8719 & 28.76 & 0.8614 & 29.19 & 0.8641 & 28.50 & 0.8569 \\
VideoINR (-m) & 29.99 & 0.8751 & 29.28 & 0.8655 & 29.68 & 0.8690 & 29.04 & 0.8606 \\
VideoINR (-s) & 29.86 & 0.8741 & 29.20 & 0.8654 & 29.42 & 0.8678 & 28.95 & 0.8613 \\
\hline
\end{tabular}
\label{tab:ablation-design}
\end{table*}

\begin{table*}[t]
\caption{\textbf{Ablation study on VideoINR trained with different data settings.} Evaluated on GOPRO-\textit{Average}. \textit{-$\times4$} refers to fixing the down-sampling space scale to $\times4$ throughout the training. \textit{-continuous} refers to training VideoINR by continuous space scales from scratch.}
\vspace{-0.75mm}
\centering
\begin{tabular}{c|cccccc|cccc}
\hline
\multirow{2}{*}{\begin{tabular}[c]{@{}c@{}}Training\\Settings\end{tabular}} & \multicolumn{2}{c}{Space $\times$2} & \multicolumn{2}{c}{Space $\times$3} & \multicolumn{2}{c|}{Space $\times$4} & \multicolumn{2}{c}{Space $\times$6} & \multicolumn{2}{c}{Space $\times$12} \\ 
 & PSNR & SSIM & PSNR & SSIM & PSNR & SSIM & PSNR & SSIM & PSNR & SSIM\\
\hline \hline
VideoINR & 29.61 & 0.8734 & 29.14 & 0.8685 & 29.41 & 0.8669 & 25.40 & 0.7590 & 24.11 & 0.6913 \\
VideoINR (\textit{-$\times4$}) & 28.25 & 0.8490 & 28.62 & 0.8626 & 29.50 & 0.8696 & 25.24 & 0.7567 & 23.82 & 0.6857 \\
VideoINR (\textit{-continuous}) & 27.46 & 0.8268 & 28.35 & 0.8507 & 28.82 & 0.8541 & 25.10 & 0.7533 & 23.62 & 0.6801   \\
\hline
\end{tabular}
\label{tab:ablation-data}
\vspace{-2mm}
\end{table*}

\noindent\textbf{Qualitative Results} We demonstrate a qualitative comparison in Figure~\ref{fig:multiframe}. We compare VideoINR with two STVSR methods, DAIN + BasicVSR and TMNet. The selected temporal coordinates of the first sample are in the training distribution, while the coordinates of the second sample are out-of-distribution. We find that the performance of DAIN + BasicVSR degrades in out-of-distribution circumstances (see the rider's head in the second sample). TMNet fails to recover objects with large motion between two input frames (see the flowers in the first sample). The performance of VideoINR is steady across both in-distribution and out-of-distribution temporal coordinates, indicating that learning continuous video representations helps to improve model generalization in STVSR task.

\subsection{Ablation Study}
\label{sec:ablation}
\noindent\textbf{Motion Flow Field. }Motion flow is one critical component of VideoINR. Previous video interpolation methods~\cite{jiang2018super,huang2020rife} have already demonstrated that such a learnable flow helps to interpolate frames with sharp edges and clear details. We propose that the motion flow field brings two main advantages. First, the flow field could capture non-local information and temporal contexts of large motions.  Second, we explicitly apply spatial warping on features, which works as an inductive bias for the training. In Table~\ref{tab:ablation-design} between VideoINR and VideoINR (-f), we show that the performance degrades when the motion flow is not incorporated.

\noindent\textbf{VideoINR trained with different data settings.} In Table~\ref{tab:ablation-data}, we compare the performances of VideoINR trained on different data settings. As noted before, VideoINR follows a two-stage training strategy: fixed down-sampling space scale for the first stage and continuous space scales sampled from a uniform distribution for the second stage. VideoINR-$\times4$ indicates that the space scale is fixed to $\times4$ throughout the training of VideoINR. VideoINR-$\textit{continuous}$ represents VideoINR trained with continuous down-sampling space scales from scratch. We find that the performance suffers a significant drop when we train VideoINR only on continuous scales. We hypothesize this is because the network needs to learn spatial and temporal representations at the same time, and it becomes extremely difficult to learn such temporal representation when the scale of spatial features keeps varying. Besides, we observe that training VideoINR with a fixed space scale achieves slightly better performance for that specific scale. However, its generalization performance is competed by VideoINR trained by two stages, which is demonstrated by the comparisons between VideoINR and VideoINR (\textit{-$\times4$}) on space scales other than $\times4$.

\noindent\textbf{Other design choices.} We provide more ablation studies in Table~\ref{tab:ablation-design}. By comparing VideoINR with VideoINR (-m), we find that the proposed multi-scale feature aggregation contributes to performance improvement. We also try to replace SpatialINR and TemporalINR by a single network, that is, we use one network only for generating the continuous motion flow, and apply spatial warping only on the encoded feature and input frames. The results between VideoINR and VideoINR (-s) indicate that using two functions for representing space and time outperforms only one network for them all.

\section{Discussion}
\label{sec:conclusion}

\noindent\textbf{Conclusion.} In this paper, we present Video Implicit Neural Representation (VideoINR). It can represent videos in arbitrary spatial and temporal resolution, which brings natural advantages for solving Space-Time Video Super-Resolution (STVSR) tasks. Extensive experiments show that VideoINR performs competitively with state-of-the-art STVSR methods on common up-sampling scales and outperforms prior works by a large margin on out-of-distribution scales. 

\noindent\textbf{Limitations and Future Work.} We observe that there exist few cases for which VideoINR does not perform very well. These cases typically need to handle very large motions, which is still an open challenge for video interpolation.

\vspace{1em}
{\textbf{Acknowledgements.}~This work was supported, in part, by gifts from Picsart.}

{\small
\bibliographystyle{ieee_fullname}
\bibliography{main}

\begin{thebibliography}{10}\itemsep=-1pt

\bibitem{anokhin2021image}
Ivan Anokhin, Kirill Demochkin, Taras Khakhulin, Gleb Sterkin, Victor
  Lempitsky, and Denis Korzhenkov.
\newblock Image generators with conditionally-independent pixel synthesis.
\newblock In {\em Proceedings of the IEEE/CVF Conference on Computer Vision and
  Pattern Recognition}, pages 14278--14287, 2021.

\bibitem{bao2019dain}
Wenbo Bao, Wei-Sheng Lai, Chao Ma, Xiaoyun Zhang, Zhiyong Gao, and Ming-Hsuan
  Yang.
\newblock Depth-aware video frame interpolation.
\newblock In {\em Proceedings of the IEEE/CVF Conference on Computer Vision and
  Pattern Recognition}, pages 3703--3712, 2019.

\bibitem{caballero2017vsr2}
Jose Caballero, Christian Ledig, Andrew Aitken, Alejandro Acosta, Johannes
  Totz, Zehan Wang, and Wenzhe Shi.
\newblock Real-time video super-resolution with spatio-temporal networks and
  motion compensation.
\newblock In {\em Proceedings of the IEEE/CVF Conference on Computer Vision and
  Pattern Recognition}, pages 4778--4787, 2017.

\bibitem{carreira2017quo}
Joao Carreira and Andrew Zisserman.
\newblock Quo vadis, action recognition? a new model and the kinetics dataset.
\newblock In {\em proceedings of the IEEE Conference on Computer Vision and
  Pattern Recognition}, pages 6299--6308, 2017.

\bibitem{chan2021pi}
Eric~R Chan, Marco Monteiro, Petr Kellnhofer, Jiajun Wu, and Gordon Wetzstein.
\newblock pi-gan: Periodic implicit generative adversarial networks for
  3d-aware image synthesis.
\newblock In {\em Proceedings of the IEEE/CVF Conference on Computer Vision and
  Pattern Recognition}, pages 5799--5809, 2021.

\bibitem{chan2021basicvsr}
Kelvin~CK Chan, Xintao Wang, Ke Yu, Chao Dong, and Chen~Change Loy.
\newblock Basicvsr: The search for essential components in video
  super-resolution and beyond.
\newblock In {\em Proceedings of the IEEE/CVF Conference on Computer Vision and
  Pattern Recognition}, pages 4947--4956, 2021.

\bibitem{chen2021liif}
Yinbo Chen, Sifei Liu, and Xiaolong Wang.
\newblock Learning continuous image representation with local implicit image
  function.
\newblock In {\em Proceedings of the IEEE/CVF Conference on Computer Vision and
  Pattern Recognition}, pages 8628--8638, 2021.

\bibitem{Chen_2019_CVPR}
Zhiqin Chen and Hao Zhang.
\newblock Learning implicit fields for generative shape modeling.
\newblock In {\em Proceedings of the IEEE/CVF Conference on Computer Vision and
  Pattern Recognition (CVPR)}, June 2019.

\bibitem{dai2017deformable}
Jifeng Dai, Haozhi Qi, Yuwen Xiong, Yi Li, Guodong Zhang, Han Hu, and Yichen
  Wei.
\newblock Deformable convolutional networks.
\newblock In {\em Proceedings of the IEEE/CVF Conference on Computer Vision and
  Pattern Recognition}, pages 764--773, 2017.

\bibitem{deng2020nasa}
Boyang Deng, John~P Lewis, Timothy Jeruzalski, Gerard Pons-Moll, Geoffrey
  Hinton, Mohammad Norouzi, and Andrea Tagliasacchi.
\newblock Nasa neural articulated shape approximation.
\newblock In {\em Computer Vision--ECCV 2020: 16th European Conference,
  Glasgow, UK, August 23--28, 2020, Proceedings, Part VII 16}, pages 612--628.
  Springer, 2020.

\bibitem{devries2021unconstrained}
Terrance DeVries, Miguel~Angel Bautista, Nitish Srivastava, Graham~W Taylor,
  and Joshua~M Susskind.
\newblock Unconstrained scene generation with locally conditioned radiance
  fields.
\newblock {\em arXiv preprint arXiv:2104.00670}, 2021.

\bibitem{feichtenhofer2019slowfast}
Christoph Feichtenhofer, Haoqi Fan, Jitendra Malik, and Kaiming He.
\newblock Slowfast networks for video recognition.
\newblock In {\em Proceedings of the IEEE/CVF international conference on
  computer vision}, pages 6202--6211, 2019.

\bibitem{genova2020local}
Kyle Genova, Forrester Cole, Avneesh Sud, Aaron Sarna, and Thomas Funkhouser.
\newblock Local deep implicit functions for 3d shape.
\newblock In {\em Proceedings of the IEEE/CVF Conference on Computer Vision and
  Pattern Recognition}, pages 4857--4866, 2020.

\bibitem{genova2019learning}
Kyle Genova, Forrester Cole, Daniel Vlasic, Aaron Sarna, William~T Freeman, and
  Thomas Funkhouser.
\newblock Learning shape templates with structured implicit functions.
\newblock In {\em Proceedings of the IEEE/CVF International Conference on
  Computer Vision}, pages 7154--7164, 2019.

\bibitem{haris2020starnet}
Muhammad Haris, Greg Shakhnarovich, and Norimichi Ukita.
\newblock Space-time-aware multi-resolution video enhancement.
\newblock In {\em Proceedings of the IEEE/CVF Conference on Computer Vision and
  Pattern Recognition}, pages 2859--2868, 2020.

\bibitem{huang2020rife}
Zhewei Huang, Tianyuan Zhang, Wen Heng, Boxin Shi, and Shuchang Zhou.
\newblock Rife: Real-time intermediate flow estimation for video frame
  interpolation.
\newblock {\em arXiv preprint arXiv:2011.06294}, 2020.

\bibitem{isobe2020videovsr}
Takashi Isobe, Xu Jia, Shuhang Gu, Songjiang Li, Shengjin Wang, and Qi Tian.
\newblock Video super-resolution with recurrent structure-detail network.
\newblock In {\em European Conference on Computer Vision}, pages 645--660.
  Springer, 2020.

\bibitem{jiang2018super}
Huaizu Jiang, Deqing Sun, Varun Jampani, Ming-Hsuan Yang, Erik Learned-Miller,
  and Jan Kautz.
\newblock Super slomo: High quality estimation of multiple intermediate frames
  for video interpolation.
\newblock In {\em Proceedings of the IEEE/CVF Conference on Computer Vision and
  Pattern Recognition}, pages 9000--9008, 2018.

\bibitem{jo2018deepvsr}
Younghyun Jo, Seoung~Wug Oh, Jaeyeon Kang, and Seon~Joo Kim.
\newblock Deep video super-resolution network using dynamic upsampling filters
  without explicit motion compensation.
\newblock In {\em Proceedings of the IEEE/CVF Conference on Computer Vision and
  Pattern Recognition}, pages 3224--3232, 2018.

\bibitem{karras2021alias}
Tero Karras, Miika Aittala, Samuli Laine, Erik H{\"a}rk{\"o}nen, Janne
  Hellsten, Jaakko Lehtinen, and Timo Aila.
\newblock Alias-free generative adversarial networks.
\newblock {\em arXiv preprint arXiv:2106.12423}, 2021.

\bibitem{kim2020fisr}
Soo~Ye Kim, Jihyong Oh, and Munchurl Kim.
\newblock Fisr: deep joint frame interpolation and super-resolution with a
  multi-scale temporal loss.
\newblock In {\em Proceedings of the AAAI Conference on Artificial
  Intelligence}, volume~34, pages 11278--11286, 2020.

\bibitem{kingma2014adam}
Diederik~P Kingma and Jimmy Ba.
\newblock Adam: A method for stochastic optimization.
\newblock {\em arXiv preprint arXiv:1412.6980}, 2014.

\bibitem{lai2017char}
Wei-Sheng Lai, Jia-Bin Huang, Narendra Ahuja, and Ming-Hsuan Yang.
\newblock Deep laplacian pyramid networks for fast and accurate
  super-resolution.
\newblock In {\em Proceedings of the IEEE conference on computer vision and
  pattern recognition}, pages 624--632, 2017.

\bibitem{liu2011vid4}
Ce Liu and Deqing Sun.
\newblock A bayesian approach to adaptive video super resolution.
\newblock In {\em Proceedings of the IEEE/CVF Conference on Computer Vision and
  Pattern Recognition}, pages 209--216. IEEE, 2011.

\bibitem{Mescheder_2019_CVPR}
Lars Mescheder, Michael Oechsle, Michael Niemeyer, Sebastian Nowozin, and
  Andreas Geiger.
\newblock Occupancy networks: Learning 3d reconstruction in function space.
\newblock In {\em Proceedings of the IEEE/CVF Conference on Computer Vision and
  Pattern Recognition (CVPR)}, June 2019.

\bibitem{meyer2015phasevfi}
Simone Meyer, Oliver Wang, Henning Zimmer, Max Grosse, and Alexander
  Sorkine-Hornung.
\newblock Phase-based frame interpolation for video.
\newblock In {\em Proceedings of the IEEE/CVF Conference on Computer Vision and
  Pattern Recognition}, pages 1410--1418, 2015.

\bibitem{michalkiewicz2019implicit}
Mateusz Michalkiewicz, Jhony~K Pontes, Dominic Jack, Mahsa Baktashmotlagh, and
  Anders Eriksson.
\newblock Implicit surface representations as layers in neural networks.
\newblock In {\em Proceedings of the IEEE/CVF International Conference on
  Computer Vision}, pages 4743--4752, 2019.

\bibitem{mildenhall2020nerf}
Ben Mildenhall, Pratul~P Srinivasan, Matthew Tancik, Jonathan~T Barron, Ravi
  Ramamoorthi, and Ren Ng.
\newblock Nerf: Representing scenes as neural radiance fields for view
  synthesis.
\newblock In {\em European conference on computer vision}, pages 405--421.
  Springer, 2020.

\bibitem{mudenagudi2010stvsr}
Uma Mudenagudi, Subhashis Banerjee, and Prem~Kumar Kalra.
\newblock Space-time super-resolution using graph-cut optimization.
\newblock {\em IEEE Transactions on Pattern Analysis and Machine Intelligence},
  33(5):995--1008, 2010.

\bibitem{nah2017gopro}
Seungjun Nah, Tae Hyun~Kim, and Kyoung Mu~Lee.
\newblock Deep multi-scale convolutional neural network for dynamic scene
  deblurring.
\newblock In {\em Proceedings of the IEEE/CVF Conference on Computer Vision and
  Pattern Recognition}, pages 3883--3891, 2017.

\bibitem{niklaus2018kernelvfi3}
Simon Niklaus and Feng Liu.
\newblock Context-aware synthesis for video frame interpolation.
\newblock In {\em Proceedings of the IEEE/CVF Conference on Computer Vision and
  Pattern Recognition}, pages 1701--1710, 2018.

\bibitem{niklaus2020softmax}
Simon Niklaus and Feng Liu.
\newblock Softmax splatting for video frame interpolation.
\newblock In {\em Proceedings of the IEEE/CVF Conference on Computer Vision and
  Pattern Recognition}, pages 5437--5446, 2020.

\bibitem{niklaus2017kernelvfi1}
Simon Niklaus, Long Mai, and Feng Liu.
\newblock Video frame interpolation via adaptive convolution.
\newblock In {\em Proceedings of the IEEE/CVF Conference on Computer Vision and
  Pattern Recognition}, pages 670--679, 2017.

\bibitem{niklaus2017kernelvfi2}
Simon Niklaus, Long Mai, and Feng Liu.
\newblock Video frame interpolation via adaptive separable convolution.
\newblock In {\em Proceedings of the IEEE International Conference on Computer
  Vision}, pages 261--270, 2017.

\bibitem{Park_2019_CVPR}
Jeong~Joon Park, Peter Florence, Julian Straub, Richard Newcombe, and Steven
  Lovegrove.
\newblock Deepsdf: Learning continuous signed distance functions for shape
  representation.
\newblock In {\em Proceedings of the IEEE/CVF Conference on Computer Vision and
  Pattern Recognition (CVPR)}, June 2019.

\bibitem{reda2019cyclevfi}
Fitsum~A Reda, Deqing Sun, Aysegul Dundar, Mohammad Shoeybi, Guilin Liu,
  Kevin~J Shih, Andrew Tao, Jan Kautz, and Bryan Catanzaro.
\newblock Unsupervised video interpolation using cycle consistency.
\newblock In {\em Proceedings of the IEEE/CVF International Conference on
  Computer Vision}, pages 892--900, 2019.

\bibitem{schwarz2020graf}
Katja Schwarz, Yiyi Liao, Michael Niemeyer, and Andreas Geiger.
\newblock Graf: Generative radiance fields for 3d-aware image synthesis.
\newblock {\em arXiv preprint arXiv:2007.02442}, 2020.

\bibitem{shahar2011stvsr}
Oded Shahar, Alon Faktor, and Michal Irani.
\newblock {\em Space-time super-resolution from a single video}.
\newblock IEEE, 2011.

\bibitem{shechtman2002stvsr}
Eli Shechtman, Yaron Caspi, and Michal Irani.
\newblock Increasing space-time resolution in video.
\newblock In {\em European Conference on Computer Vision}, pages 753--768.
  Springer, 2002.

\bibitem{sitzmann2020siren}
Vincent Sitzmann, Julien Martel, Alexander Bergman, David Lindell, and Gordon
  Wetzstein.
\newblock Implicit neural representations with periodic activation functions.
\newblock {\em Advances in Neural Information Processing Systems}, 33, 2020.

\bibitem{skorokhodov2021adversarial}
Ivan Skorokhodov, Savva Ignatyev, and Mohamed Elhoseiny.
\newblock Adversarial generation of continuous images.
\newblock In {\em Proceedings of the IEEE/CVF Conference on Computer Vision and
  Pattern Recognition}, pages 10753--10764, 2021.

\bibitem{su2017adobe}
Shuochen Su, Mauricio Delbracio, Jue Wang, Guillermo Sapiro, Wolfgang Heidrich,
  and Oliver Wang.
\newblock Deep video deblurring for hand-held cameras.
\newblock In {\em Proceedings of the IEEE/CVF Conference on Computer Vision and
  Pattern Recognition}, pages 1279--1288, 2017.

\bibitem{sun2018pwcnet}
Deqing Sun, Xiaodong Yang, Ming-Yu Liu, and Jan Kautz.
\newblock Pwc-net: Cnns for optical flow using pyramid, warping, and cost
  volume.
\newblock In {\em Proceedings of the IEEE/CVF Conference on Computer Vision and
  Pattern Recognition}, pages 8934--8943, 2018.

\bibitem{tao2017vsr}
Xin Tao, Hongyun Gao, Renjie Liao, Jue Wang, and Jiaya Jia.
\newblock Detail-revealing deep video super-resolution.
\newblock In {\em Proceedings of the IEEE/CVF International Conference on
  Computer Vision}, pages 4472--4480, 2017.

\bibitem{tian2020tdan}
Yapeng Tian, Yulun Zhang, Yun Fu, and Chenliang Xu.
\newblock Tdan: Temporally-deformable alignment network for video
  super-resolution.
\newblock In {\em Proceedings of the IEEE/CVF Conference on Computer Vision and
  Pattern Recognition}, pages 3360--3369, 2020.

\bibitem{wang2019edvr}
Xintao Wang, Kelvin~CK Chan, Ke Yu, Chao Dong, and Chen Change~Loy.
\newblock Edvr: Video restoration with enhanced deformable convolutional
  networks.
\newblock In {\em Proceedings of the IEEE/CVF Conference on Computer Vision and
  Pattern Recognition Workshops}, pages 0--0, 2019.

\bibitem{wang2004ssim}
Zhou Wang, Alan~C Bovik, Hamid~R Sheikh, and Eero~P Simoncelli.
\newblock Image quality assessment: from error visibility to structural
  similarity.
\newblock {\em IEEE transactions on image processing}, 13(4):600--612, 2004.

\bibitem{xiang2020zooming}
Xiaoyu Xiang, Yapeng Tian, Yulun Zhang, Yun Fu, Jan~P Allebach, and Chenliang
  Xu.
\newblock Zooming slow-mo: Fast and accurate one-stage space-time video
  super-resolution.
\newblock In {\em Proceedings of the IEEE/CVF conference on computer vision and
  pattern recognition}, pages 3370--3379, 2020.

\bibitem{xu2021tmnet}
Gang Xu, Jun Xu, Zhen Li, Liang Wang, Xing Sun, and Ming-Ming Cheng.
\newblock Temporal modulation network for controllable space-time video
  super-resolution.
\newblock In {\em Proceedings of the IEEE/CVF Conference on Computer Vision and
  Pattern Recognition}, pages 6388--6397, 2021.

\bibitem{xu2019qvi}
Xiangyu Xu, Li Siyao, Wenxiu Sun, Qian Yin, and Ming-Hsuan Yang.
\newblock Quadratic video interpolation.
\newblock {\em Advances in Neural Information Processing Systems},
  32:1647--1656, 2019.

\bibitem{xu2021ultrasr}
Xingqian Xu, Zhangyang Wang, and Humphrey Shi.
\newblock Ultrasr: Spatial encoding is a missing key for implicit image
  function-based arbitrary-scale super-resolution.
\newblock {\em arXiv preprint arXiv:2103.12716}, 2021.

\bibitem{xue2019vimeo}
Tianfan Xue, Baian Chen, Jiajun Wu, Donglai Wei, and William~T Freeman.
\newblock Video enhancement with task-oriented flow.
\newblock {\em International Journal of Computer Vision}, 127(8):1106--1125,
  2019.

\bibitem{zhu2019deformable}
Xizhou Zhu, Han Hu, Stephen Lin, and Jifeng Dai.
\newblock Deformable convnets v2: More deformable, better results.
\newblock In {\em Proceedings of the IEEE/CVF Conference on Computer Vision and
  Pattern Recognition}, pages 9308--9316, 2019.

\bibitem{zhu2017flow}
Xizhou Zhu, Yujie Wang, Jifeng Dai, Lu Yuan, and Yichen Wei.
\newblock Flow-guided feature aggregation for video object detection.
\newblock In {\em Proceedings of the IEEE International Conference on Computer
  Vision}, pages 408--417, 2017.

\end{thebibliography}
}
\clearpage
\appendix

\section{Implementation Details}
All models included in experiments are trained from scratch to perform fair comparisons. 
For video frame interpolation methods incorporated in the experiments (\ie Super SloMo~\cite{jiang2018super}, QVI~\cite{xu2019qvi}, and DAIN~\cite{bao2019dain}), we train them on Adobe240 dataset~\cite{su2017adobe}. We keep all the training settings the same as proposed in their original papers, including the optimizer, initial learning rate, learning rate decay strategy, and the number of training epochs.  For data settings, $9$ consecutive frames are selected from video clips for training in every iteration. Networks take the first and last frames as inputs and generate intermediate $7$ frames. We calculate loss between generated frames and the original ground-truth frames. Each video frame is resized to have a shorter spatial dimension of 360, and a random crop of 352$\times$352 is performed. 

Zooming SlowMo~\cite{xiang2020zooming} and TMNet~\cite{xu2021tmnet} are two STVSR models included in our experiments. Zooming SlowMo only supports fixed frame interpolation, and the interpolation time is set to $0$, $0.5$, $1$ in the original paper. Following their settings, we also train the model from scratch to interpolate the fixed time instances. To ensure that the input video frames of all models are of the same frame rate, we extract $9$ consecutive frames from video clips and take the $1^{st}$ and $9^{th}$ frames as inputs. We then down-sample the input frames via Bicubic interpolation by a factor of 4 and use the network to predict the high-resolution versions of the $1^{st}$, $5^{th}$ and $9^{th}$ frames. TMNet supports arbitrary frame interpolation. In its paper, the authors mention that TMNet needs a two-stage training process for convergence, and we follow their suggestions. In the first stage, we pre-train the network on the Vimeo90K dataset~\cite{xue2019vimeo}. The Vimeo90K dataset consists of 7-frame video sequences. We use the $1^{st}$, $3^{rd}$, $5^{th}$, and $7^{th}$ frames after down-sampling as the network inputs and predict the high-resolution results of all the $7$ frames, which means that the interpolation time is set to $0$, $0.5$, $1$ in this stage. In the second stage, we select $9$ consecutive frames from video clips, and the $1^{st}$ and $9^{th}$ frames are taken as inputs. After down-sampling, we use the network to generate high-resolution predictions of all $9$ frames and calculate the loss value with the original high-resolution frames. TMNet is trained with more data, which may lead to advantages in the experiments.

For the training of VideoINR, we select $9$ consecutive frames and down-sample the $1^{st}$ and $9^{th}$ frames as model inputs. In each iteration, We randomly select three frames from the 9-frame video sequence and use the network to generate high-resolution predictions at the time instances of the three selected frames. 

We keep the training settings unchanged for Zooming SlowMo, TMNet, and VideoINR. All three models are optimized with the Charbonnier loss function~\cite{lai2017char}.

\begin{figure}
    \centering
    \includegraphics[width=\linewidth]{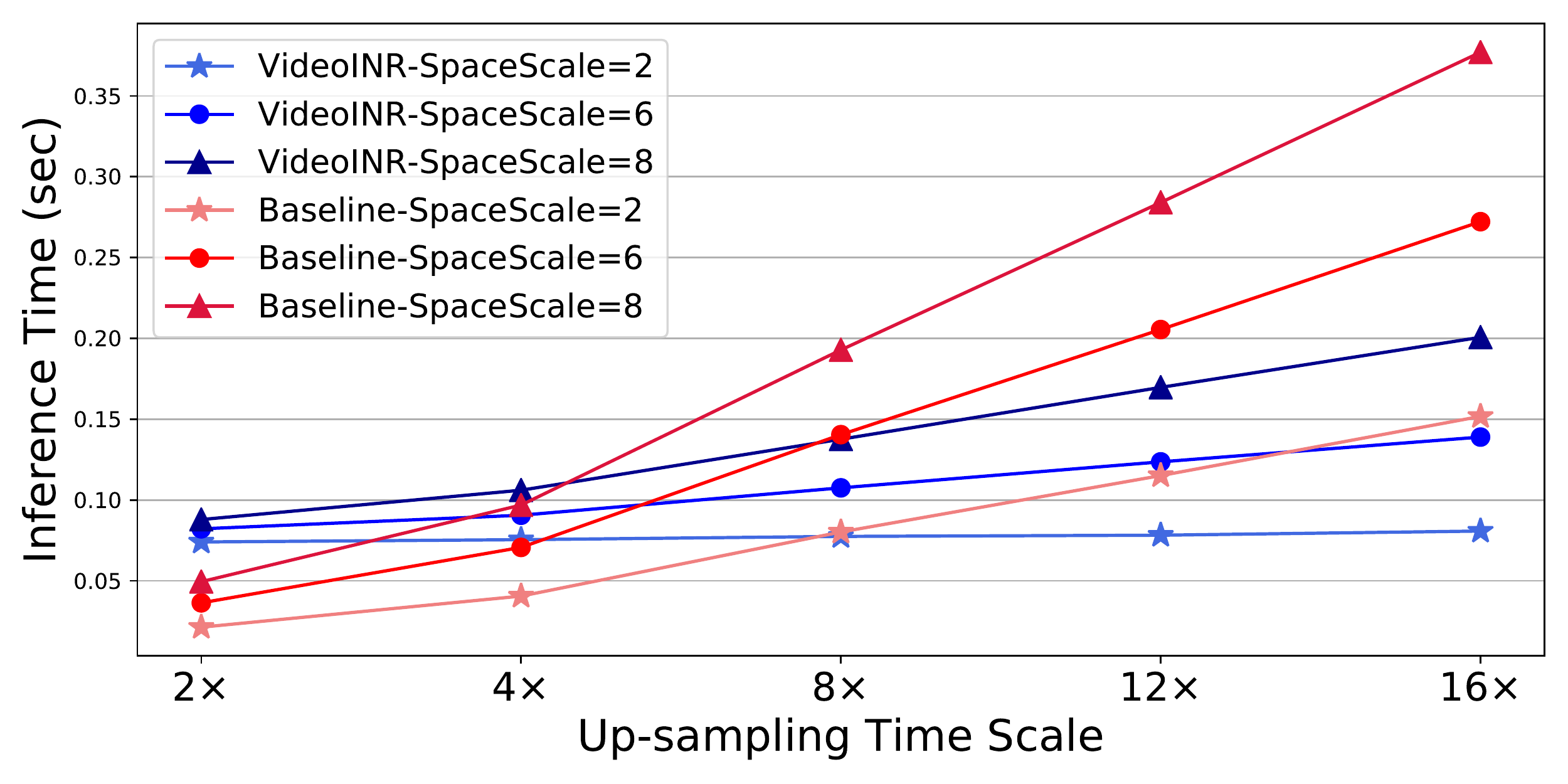}
    \vspace{-2.0em}
    \caption{Inference time comparisons on different space scales.}
    \label{fig:time_supp}
\end{figure}

\begin{figure}
    \centering
    \includegraphics[width=\linewidth]{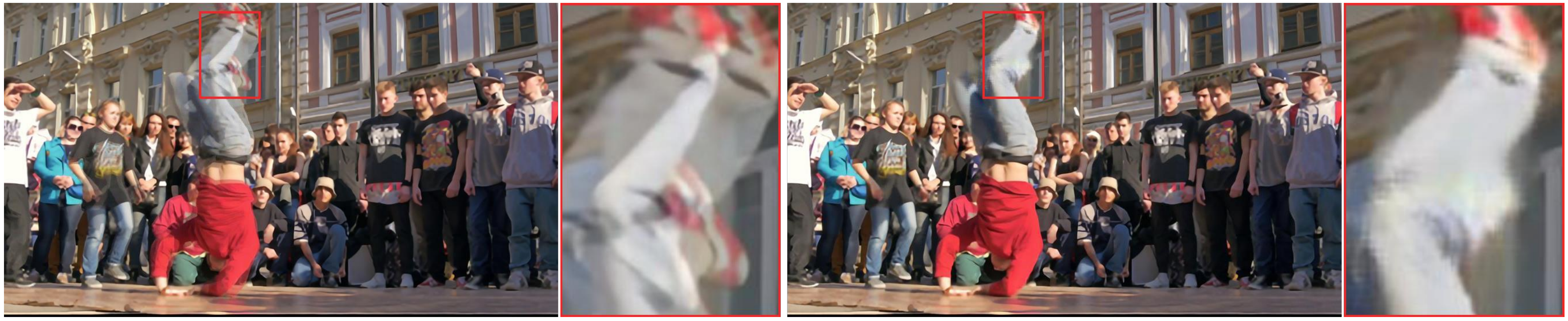}
    \vspace{-1.5em}
    \caption{Failure case. Left is the overlay of frames at t=0 and t=1. Right is the interpolated frame at t=0.5}
    \vspace{-1.0em}
    \label{fig:fail}
\end{figure}

\section{Efficiency on Different Scales}
To evaluate the efficiency of VideoINR on different up-sampling space scales, we provide more inference time comparisons in Figure~\ref{fig:time_supp}. We select the two-stage method composed of SuperSlomo and LIIF as the baseline, as it supports arbitrary up-sampling scales on both space and time.

\section{Limitations}
In some challenging cases, large motion and occlusion result in errors on the motion flow field, leading to blurred results with unclear boundaries. We show a failure cases of VideoINR in Figure~\ref{fig:fail}.

\section{Additional Qualitative Results}
\begin{figure*}
    \centering
    \vspace{-0.3em}
    \includegraphics[width=1.05\linewidth]{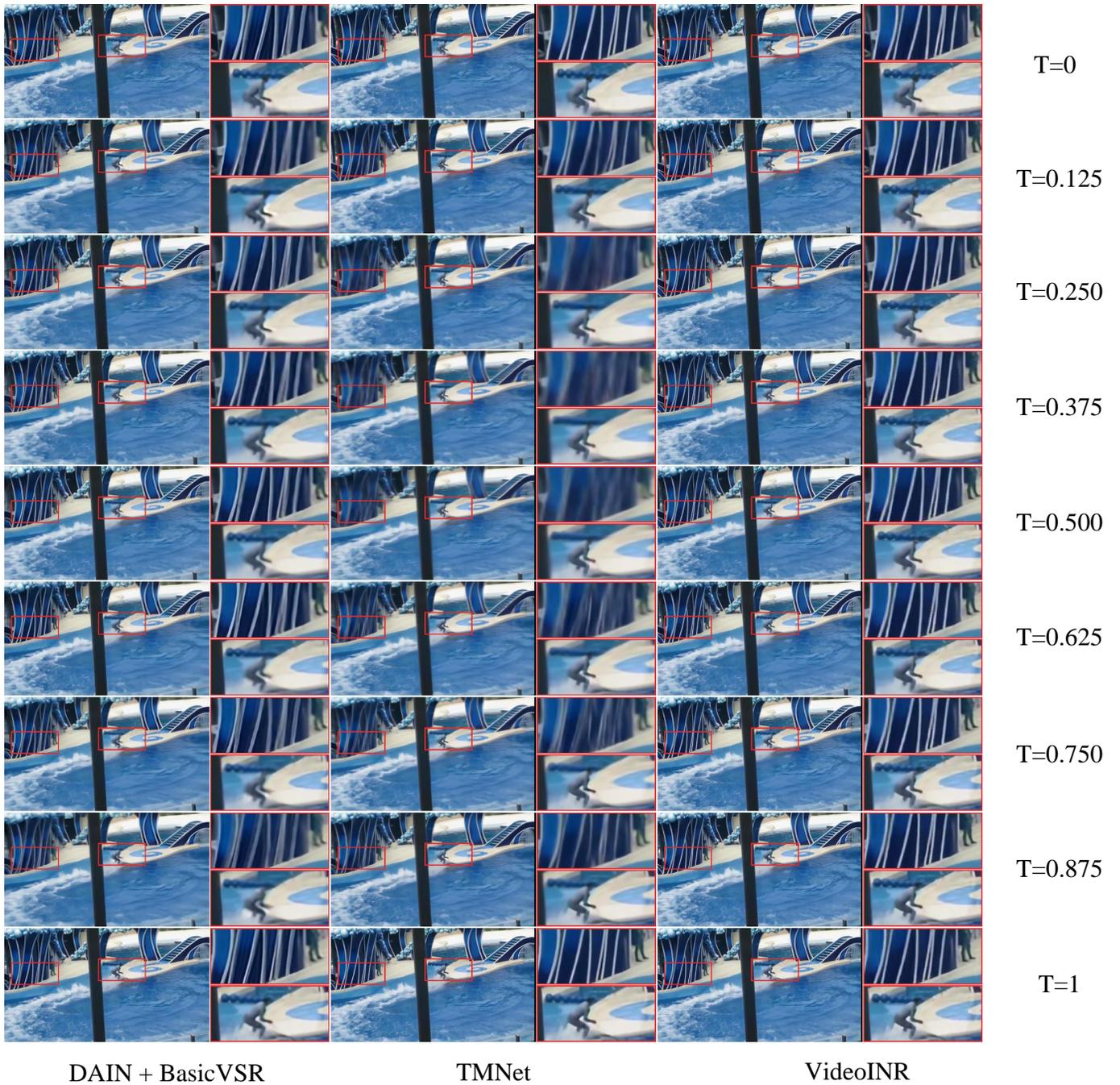}
    \caption{Qualitative comparisons of different STVSR methods on in-distribution time scale. Best zoom in for better visualization.}
    \vspace{-1em}
    \label{fig:1}
\end{figure*}

\begin{figure*}
    \centering
    \vspace{-0.3em}
    \includegraphics[width=1.05\linewidth]{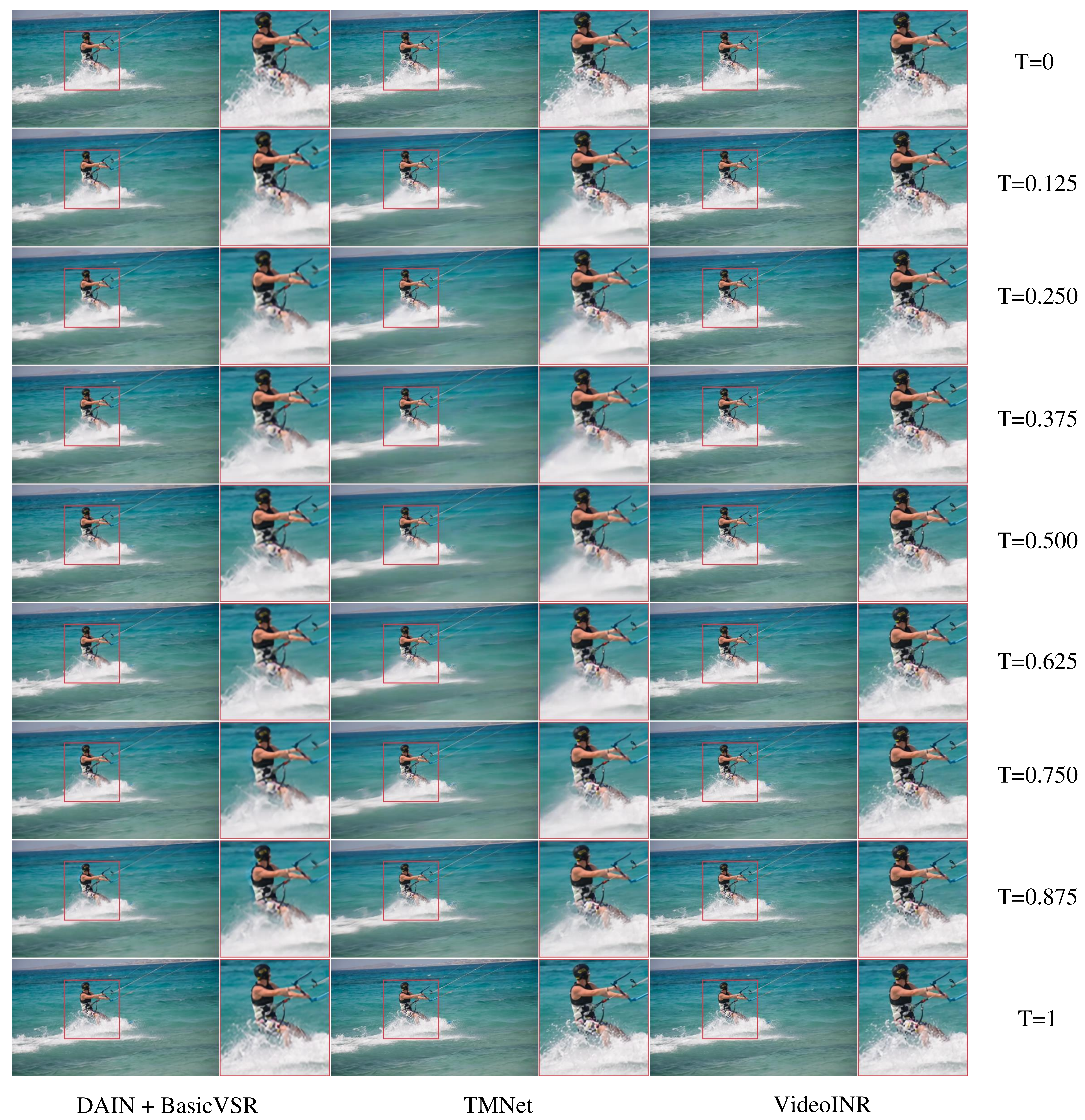}
    \caption{Qualitative comparisons of different STVSR methods on in-distribution time scale. Best zoom in for better visualization.}
    \vspace{-1em}
    \label{fig:2}
\end{figure*}

\begin{figure*}
    \centering
    \vspace{-0.3em}
    \includegraphics[width=1.05\linewidth]{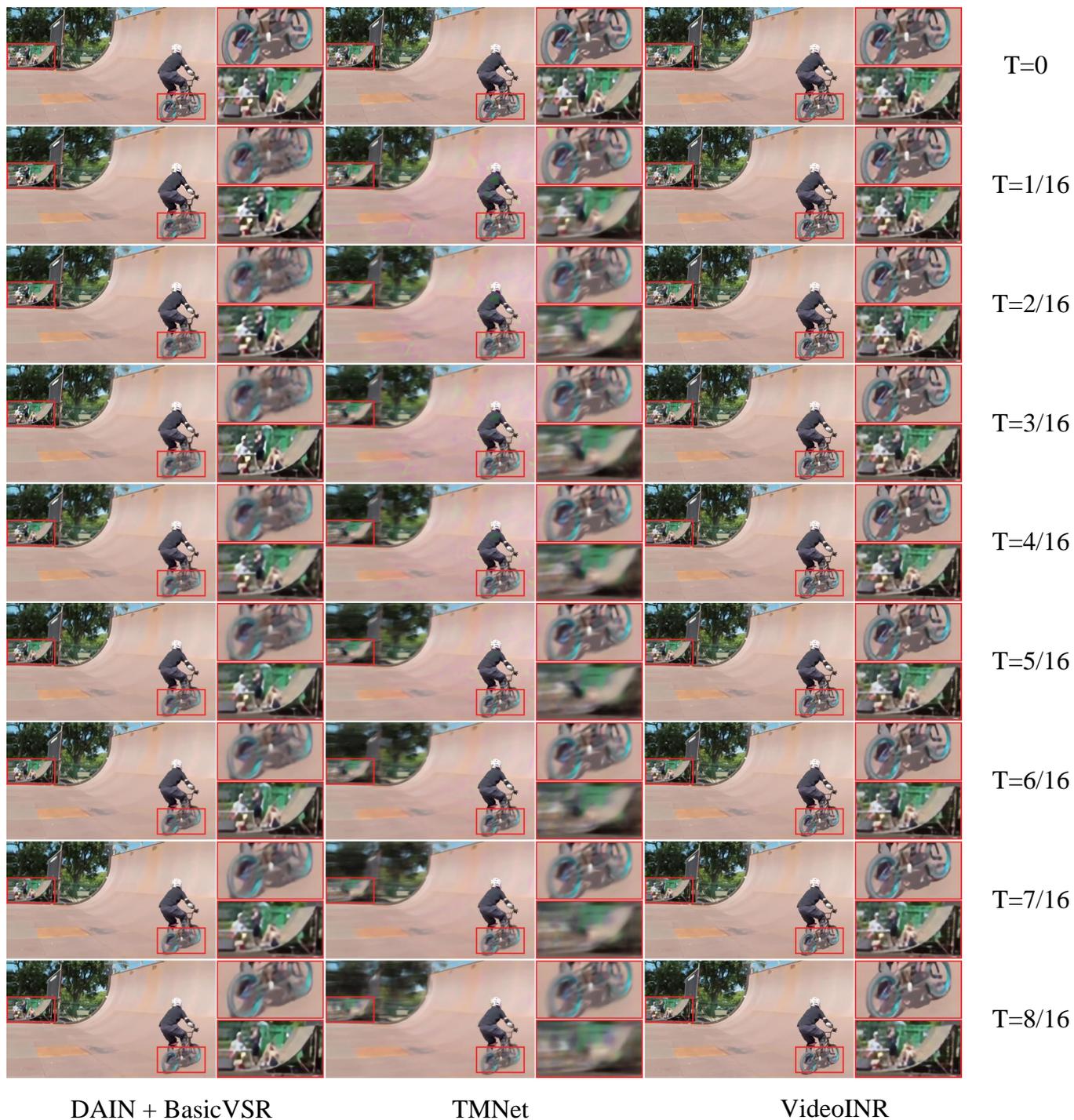}
    \caption{Qualitative comparisons of different STVSR methods on out-of-distribution time scale. Best zoom in for better visualization.}
    \vspace{-1em}
    \label{fig:3}
\end{figure*}

\begin{figure*}
    \centering
    \vspace{-0.3em}
    \includegraphics[width=1.05\linewidth]{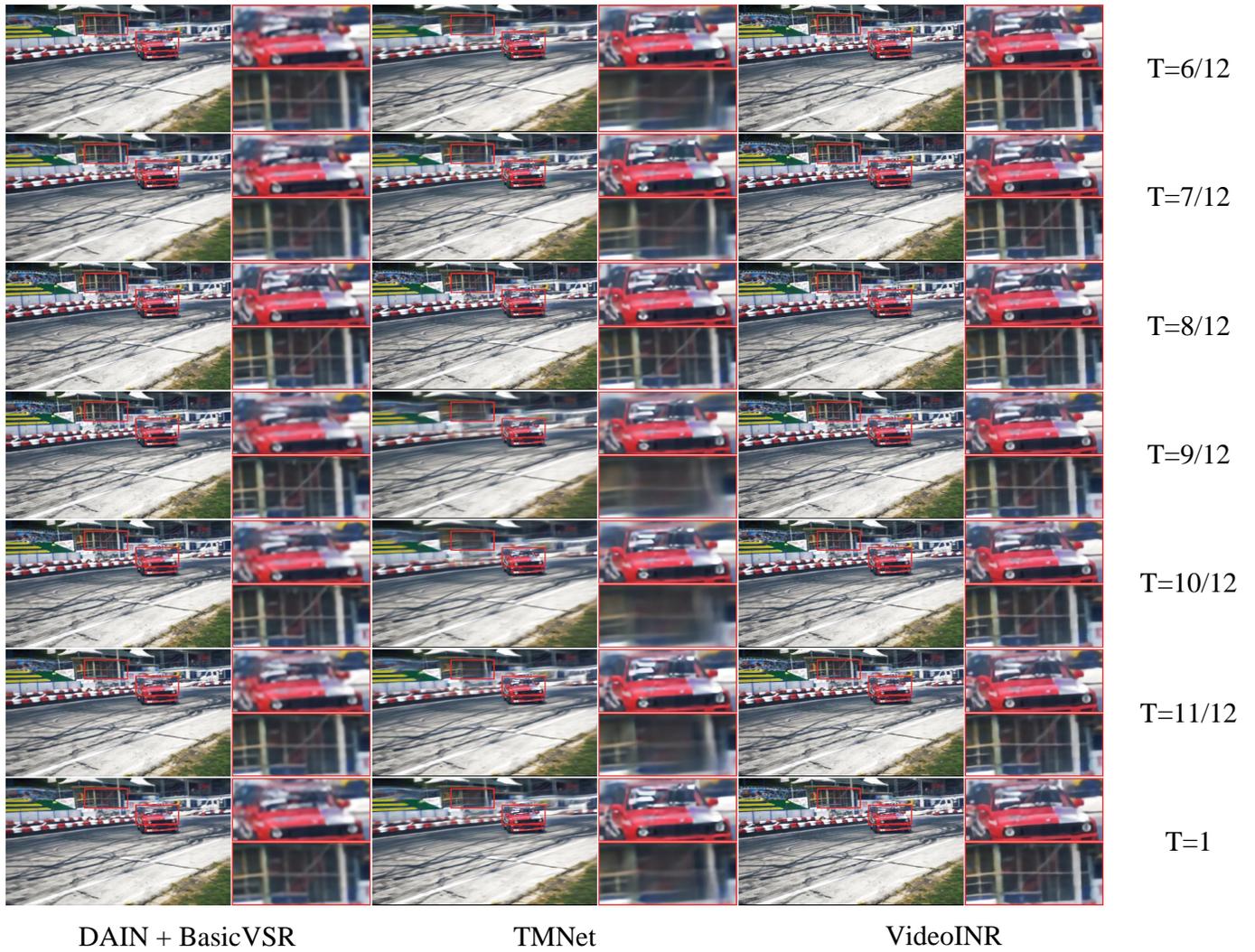}
    \caption{Qualitative comparisons of different STVSR methods on out-of-distribution time scale. Best zoom in for better visualization.}
    \vspace{-1em}
    \label{fig:4}
\end{figure*}

We provide more qualitative results in Figures~\ref{fig:1},\ref{fig:2},\ref{fig:3},\ref{fig:4}. We compare VideoINR with two STVSR methods, DAIN~\cite{bao2019dain} + BasicVSR~\cite{chan2021basicvsr} and TMNet~\cite{xu2021tmnet}. The up-sampling space scale is set to 4 for all examples. In Figure~\ref{fig:1}, \ref{fig:2}, we set the time scale for interpolation to 8, which is in our training distribution. We observe that DAIN + BasicVSR and TMNet tend to generate blurry regions or artifacts. In contrast, the results of VideoINR are consistent and aligned across two input frames, with sharp edges and clear details. In Figure~\ref{fig:3}, \ref{fig:4}, we set the time scale to 12 and 16, which are out of the training distribution. We find that VideoINR well recovers objects with large motion and preserves better textural information compared with other methods. In summary, VideoINR shows the advantages of learning continuous representation for videos, and address the Space-Time Video Super-Resolution task. More visualization results can be found in the provided video.

\end{document}